\documentclass[12pt,thmsa,a4paper]{article}
\usepackage{sw20elba}

\input tcilatex
\begin{document}

\title{\textbf{Principles of Discrete Time Mechanics:}\\
$\mathbf{II}$\textbf{. Classical Field Theory}}
\author{George Jaroszkiewicz and Keith Norton \\
Department of Mathematics, University of Nottingham\\
University Park, Nottingham NG7 2RD, UK}
\date{19$^{th}$ December 1996}
\maketitle

\begin{abstract}
\textit{We apply the principles discussed in an earlier paper to the
construction of discrete time field theories. We derive the discrete time
field equations of motion and Noether's theorem and apply them to the
Schr\"{o}dinger equation to illustrate the methodology. Stationary solutions
to the discrete time Schr\"{o}dinger wave equation are found to be identical
to standard energy eigenvalue solutions except for a fundamental limit on
the energy. Then we apply the formalism to the free neutral Klein Gordon
system, deriving the equations of motion and conserved quantities such as
the linear momentum and angular momentum. We show that there is an upper
bound on the magnitude of linear momentum for physical particle-like
solutions. We extend the formalism to the charged scalar field coupled to
Maxwell's electrodynamics in a gauge invariant way. We apply the formalism
to include the Maxwell and Dirac fields, setting the scene for second
quantisation of discrete time mechanics and discrete time Quantum
Electrodynamics.}
\end{abstract}

\section{Introduction}

THIS paper follows an earlier paper on the principles of discrete time
mechanics as formulated for point particle theories, hereafter referred to
as \textit{Paper I }\cite{J&N}. These principles are applied in this paper
to systems described by fields $\varphi ^\alpha (t,\mathbf{x})$ in one time
and $3$ spatial dimensions, where $\alpha $ denotes spin component and field
types. Units will normally be taken to give $c=\hbar =1.$ The summation
convention will apply only to repeated small Greek indices except where
otherwise stated.

In discrete time mechanics dynamical variables take on values at times $%
t_n,\;n=0,1,2,...N,$ with the discrete time intervals $\Delta t_n\equiv
t_{n+1}-t_n,$ $n=0,1,2,...$ given by $\Delta t_n=T$ , where $T>0$ is the
fundamental time unit of the system. An important principle of our
formulation is that the mechanics is regarded as exact regardless of the
magnitude of $T,$ and not an exercise in approximation. This means that
invariants of the motion are constructed to be conserved precisely and not
up to some given order in $T.$

As in \textit{Paper I} our discrete time approach takes its cues from
continuous time mechanics, but ultimately is more than an approximation to
such a formalism. In general the continuous time Lagrangians $L$ for the
systems of interest to us in this paper are the spatial integrals of
Lagrange densities $\mathcal{L}$ of the form $\mathcal{L}=\mathcal{L}(%
\mathbf{\varphi },\partial _t\mathbf{\varphi },\nabla \mathbf{\varphi })$.

Important concepts in our methodology are those of the \textit{system
function} and \textit{virtual path. }In our formalism the system function
has the status occupied by the Lagrangian in continuous time mechanics. We
may calculate equations of motion and conserved quantities if we know the
system function. We may also use the system function to define conjugate
momenta and to quantise our theory. We construct the system function from
the continuous time Lagrangian by integrating over certain trajectories,
which we refer to as virtual paths. These are not solutions to the normal
Euler-Lagrange equations of motion but paths chosen according to specific
criteria such as gauge invariance.

In our approach we shall discuss discrete time mechanics based on a temporal
lattice with successive points given by 
\begin{equation}
t_{n+1}-t_n=T,
\end{equation}
where $T$ is non-zero and independent of $n$. If $\varphi \left( t\right) $
is a dynamical variable depending on time we will normally define a virtual
path $\tilde{\varphi}$ from $\varphi _n\equiv \varphi \left( t_n\right) $ to 
$\varphi _{n+1}\equiv \varphi \left( t_{n+1}\right) $ in $\varphi $-value
space by the linear rule 
\begin{equation}
\tilde{\varphi}\left( t\right) \equiv \lambda \varphi _{n+1}+\bar{\lambda}%
\varphi _n,  \label{virt}
\end{equation}
where $\lambda $ runs from $0$ to $1$ and $\bar{\lambda}\equiv 1-\lambda .$
This means that over the interval $\left[ t_n,t_{n+1}\right] $ the time $t$
is related to $\lambda $ by the rule 
\begin{equation}
t=t_n+\lambda T=\lambda t_{n+1}+\bar{\lambda}t_n.  \label{virt2}
\end{equation}
This prescription is modified when we include gauge invariance, as discussed
in \S $8$.\ With the above virtual path prescription time derivatives are
naturally turned into differences, i.e. 
\begin{equation}
\frac \partial {\partial t}\tilde{\varphi}\left( t\right) =\frac{\varphi
_{n+1}-\varphi _n}T.
\end{equation}

It is possible to use a different temporal lattice. A popular choice in
recent years has been the so-called $q$-lattice, where we define $%
t_{n+1}\equiv qt_n$, where $q$ is real and not equal to unity. Then we would
define $\varphi _n\equiv \varphi \left( t_n\right) =\varphi \left(
q^nt_0\right) ,t_0\neq 0.$ The virtual path prescription on the interval $%
\left[ t_n,t_{n+1}\right] $ would be the same as $\left( \ref{virt2}\right) $
with 
\begin{equation}
t=t_n+\lambda \left( t_{n+1}-t_n\right) =\lambda t_{n+1}+\bar{\lambda}t_n,
\end{equation}
and so in the interval $\left[ t_n,t_{n+1}\right] $ the temporal derivative
would be replaced by the so-called $q$-derivative 
\begin{equation}
\frac \partial {\partial t}\tilde{\varphi}\left( t\right) =\frac{\varphi
_{n+1}-\varphi _n}{t_{n+1}-t_n}=\frac{\varphi \left( qt_n\right) -\varphi
\left( t_n\right) }{\left( q-1\right) t_n}.
\end{equation}
We may go further and accommodate the so-called symmetric derivative \cite
{KLIMEK.93} given by 
\begin{equation}
\frac{\bar{\partial}}{\partial t}\tilde{\varphi}\left( t\right) \equiv \frac{%
\varphi \left( qt_n\right) -\varphi \left( q^{-1}t_n\right) }{\left(
q-q^{-1}\right) t_n},
\end{equation}
but in this particular case the virtual temporal path (viewed in terms of
successive intervals) is not continuous everywhere. We may even generalise
in the manner of Klimek \cite{KLIMEK.93} and define $t_{n+1}\equiv \phi
\left( t_n\right) $, where $\phi $ is some chosen real valued function and
then the above prescription gives 
\begin{equation}
\frac \partial {\partial t}\tilde{\varphi}\left( t\right) =\frac{\varphi
\left( \phi \left( t_n\right) \right) -\varphi \left( t_n\right) }{\phi
\left( t_n\right) -t_n}.
\end{equation}

The regular temporal lattice is given by the function $\phi \left( t\right)
=t+T$ in Klimek's formulation. We shall use this lattice throughout this
paper, applying it to the Schr\"{o}dinger wave equation, the free neutral
scalar field, the free charged scalar field, the Maxwell fields and
potentials and to the Dirac field, and show how to couple these fields in a
gauge invariant way. The quantisation of our discrete time field theories
will be discussed in the next paper of this series, \textit{Paper III},
following the principles discussed in \textit{Paper I}. A detailed
discussion of quantum electrodynamics will be presented in \textit{Paper IV}
of this series.

The plan of this paper is as follows. First we review Hamilton's principal
function in continuous time field theory and then discuss the related
concept of system function in discrete time field theory. We derive the
field Cadzow's equations of motion and discuss the Maeda-Noether theorem for
the construction of invariants of the motion. Then we apply our formalism to
the Schr\"{o}dinger equation, as an example of a $\mathcal{DQ}$ process,
where $\mathcal{Q}$ represents the process of quantisation and $\mathcal{D}$
represents the process of discretisation of time. The resulting theory is a
classical field theory which is not the same as the result of the $\mathcal{%
QD}$ process discussed in \textit{Paper I}.

Then we turn to relativistic theories. The $\mathcal{D}$ process breaks
Lorentz invariance, but in a less emphatic fashion than space-time lattice
theories such as those discussed by Yamamoto et al. \cite
{YAMAMOTO.95A,YAMAMOTO.95B}. The systems we study here are the free neutral
and charged Klein-Gordon fields, the Maxwell potentials and their coupling
to the charged scalar field, and the Dirac equation. This sets the scene for
second quantisation, i.e. the $\mathcal{QDQ}$ process, which is discussed in 
\textit{Paper III}.

\section{Hamilton's principal function}

In this section we review some aspects of continuous time mechanics field
theory which are relevant to our method of discretisation. Following the
principles outlined in \textit{Paper I} we shall consider a dynamical field
system for which we have knowledge of the dynamical variables $\varphi
^\alpha $ at a sequence of times $t_n,n=0,1,2,...N,$ where $t_{n+1}\equiv
t_n+T.$ An important construction in continuous time mechanics is Hamilton's
principal function $S^n\left( T\right) ,$ defined as the integral 
\begin{equation}
S^n\left( T\right) \equiv \int\limits_{t_n}^{t_{n+1}}dt\int d^3\mathbf{x\,}%
\mathcal{L}  \label{int}
\end{equation}
over the classical solution to the field equations which satisfies the
boundary conditions 
\begin{equation}
\varphi _n^\alpha \left( \mathbf{x}\right) \equiv \varphi ^\alpha \left( t_n,%
\mathbf{x}\right) ,\hspace{0.5in}\varphi _{n+1}^\alpha \left( \mathbf{x}%
\right) \equiv \varphi ^\alpha \left( t_{n+1},\mathbf{x}\right) .
\end{equation}
$S^n(T)$ is a functional of $\mathbf{\varphi }_n$ and $\mathbf{\varphi }%
_{n+1}$ and is the continuous time analogue of the system function in
discrete time mechanics. If the momentum $\pi ^\alpha $ conjugate to $%
\varphi ^\alpha $ is defined by $\pi ^\alpha \equiv \frac{\partial \mathcal{L%
}}{\partial \dot{\varphi}^\alpha }$ and the Hamiltonian $H$ defined by 
\begin{equation}
H\equiv \int d^3\mathbf{x\,\{}\pi ^\alpha \dot{\varphi}^\alpha -\mathcal{L\}}
\end{equation}
then for an infinitesimal variation $\delta \varphi ^\alpha $ of the fields
we find 
\begin{equation}
\delta S^n\left( T\right) \stackunder{c}{=}\left[ -\delta tH+\int d^3\mathbf{%
x}\pi ^\alpha \delta \varphi ^\alpha \right] _{t_n}^{t_{n+1}},
\end{equation}
where the symbol $\stackunder{c}{=}$ denotes an equality holding over the
true or dynamical trajectory. From this we deduce the functional derivatives 
\begin{equation}
\frac{\delta S^n}{\delta \varphi _n^\alpha \left( \mathbf{x}\right) }=-\pi
_n^\alpha \left( \mathbf{x}\right) ,\;\;\;\;\;\frac{\delta S^n}{\delta
\varphi _{n+1}^\alpha \left( \mathbf{x}\right) }=\pi _{n+1}^\alpha \left( 
\mathbf{x}\right)   \label{aa2}
\end{equation}
and the partial derivatives 
\begin{equation}
\frac{\partial S^n}{\partial t_n}=H^n,\;\;\;\;\;\;\;\;\;\;\;\;\;\frac{%
\partial S^n}{\partial t_{n+1}}=-H^{n+1},  \label{aa3}
\end{equation}
where $H^n$ is the value of the Hamiltonian at time $t_n$.

By the process of constructing $S^n$ for each of the intervals $\left[
t_{n,}t_{n+1}\right] ,\;n=0,1,...,N-1,$ we arrive at an action integral from 
$t_0$ to $t_N$ which is equivalent to the action sum 
\begin{equation}
A^N\equiv \sum_{n=0}^{N-1}S^n.  \label{sm}
\end{equation}
If the end point field values are held fixed $\left( \text{at times }t_0%
\text{, }t_N\right) $ then this sum depends on the field values $\varphi
_n^\alpha \left( \mathbf{x}\right) $ at the intermediate time $t_n,$ $%
n=1,2,...,N-1.$ If now we vary these intermediate fields and apply a
variational principle we find the equations 
\begin{equation}
\frac \delta {\delta \varphi _n^\alpha \left( \mathbf{x}\right) }\left\{
S^{n-1}+S^n\right\} \stackunder{c}{=}0,\hspace{0.5in}0<n<N.  \label{can}
\end{equation}
These are the continuous time mechanics analogues of Cadzow's equations of
motion $\cite{CADZOW.70}$ in discrete time field theory, discussed in the
next section. The above equations $\left( \ref{can}\right) $ essentially
ensure the continuity of the conjugate momenta at the end points of the
sub-intervals.

\section{Discrete time field theory}

We turn now to the discretisation of Lagrange based field theories based on
the construction of the system function $F^n$ using the principles outlined
in \textit{Paper I}$.$ By definition, the system function is the integral 
\begin{equation}
F^n=\int_{t_n}^{t_{n+1}}dt\int d^3\mathbf{x\,}\mathcal{L}\left( \mathbf{%
\tilde{\varphi}},\partial _t\mathbf{\tilde{\varphi}},\nabla \mathbf{\tilde{%
\varphi}}\right) 
\end{equation}
with some choice of virtual path $\mathbf{\tilde{\varphi}}$ in field value
space. Once the temporal integral has been carried out the system function
is a function of the field values ${\varphi }_n^\alpha (\mathbf{x})\equiv {%
\varphi }^\alpha (t_n,\mathbf{x)}$ at times $t_n$, $t_{n+1}$ and their
spatial derivatives at those times.

Given the system function, the action sum $A^N$ is given by 
\begin{equation}
A^N=\sum_{n=0}^{N-1}F^n  \label{action}
\end{equation}
and it is this which replaces the action sum $\left( \ref{sm}\right) $ in
continuous time mechanics. We derive the equations of motion using $\left( 
\ref{action}\right) $ and a variational principle in the standard fashion $%
\cite{CADZOW.70}$. Consider infinitesimal variations of the fields with
fixed end-points, $\delta \mathbf{\varphi }_0(\mathbf{x)=}\delta \mathbf{%
\varphi }_N\mathbf{(x)=0}$. Assuming the fields fall off rapidly at spatial
infinity, we apply Cadzow's action principle to obtain the discrete time
field equations of motion 
\begin{equation}
\frac \delta {\delta \varphi _n^\alpha \left( \mathbf{x}\right) }\left\{
F^n+F^{n-1}\right\} \stackunder{c}{=}0,\;\;\;\;\;\;\;\;0<n<N,
\end{equation}
which are equivalent to 
\begin{equation}
\frac \partial {\partial \varphi _n^\alpha }\left\{ \mathcal{F}^n+\mathcal{F}%
^{n-1}\right\} \stackunder{c}{=}\nabla \mathbf{\cdot }\frac \partial
{\partial \nabla \varphi _n^\alpha }\left\{ \mathcal{F}^n+\mathcal{F}%
^{n-1}\right\} ,  \label{Cadzow}
\end{equation}
where $\mathcal{F}^n\equiv \mathcal{F}(\mathbf{\varphi }_n,\nabla \mathbf{%
\varphi }_n,\mathbf{\varphi }_{n+1},\nabla \mathbf{\varphi }_{n+1})$ is the
system function density defined by 
\begin{equation}
\mathcal{F}^n\equiv \int_{t_n}^{t_{n+1}}dt\,\mathcal{L}\left( \mathbf{\tilde{%
\varphi}},\partial _t\mathbf{\tilde{\varphi}},\nabla \mathbf{\tilde{\varphi}}%
\right) .
\end{equation}

The construction of constants of the motion is straightforward using
Noether's theorem applied to discrete systems. Consider an infinitesimal
transformation $\delta \mathbf{\varphi }_n,\delta \mathbf{\varphi }_{n+1}$
of the fields which leaves the system function unchanged. Then using the
equation of motion we deduce that the quantity 
\begin{equation}
C^n\equiv \int d^3\mathbf{x}\left\{ \frac{\partial \mathcal{F}^n}{\partial {%
\varphi }_n^\alpha }-\nabla \mathbf{\cdot }\frac{\partial \mathcal{F}^n}{%
\partial \nabla {\varphi }_n^\alpha }\right\} \delta \varphi _n^\alpha 
\end{equation}
is conserved on dynamical trajectories. In \textit{Paper I} we referred to
this as the Maeda-Noether theorem \cite{MAEDA.81}.

For linear momentum consider a transformation of the fields by the
infinitesimal shift 
\begin{equation}
\varphi _n^\alpha (\mathbf{x)\rightarrow }\varphi _n^{\alpha \prime }(%
\mathbf{x})=\varphi _n^\alpha (\mathbf{x})+\delta \mathbf{a\cdot }\nabla
\varphi _n^\alpha (\mathbf{x}).
\end{equation}
Assuming that the Lagrange density is not explicitly dependent on position
then we find that the conserved linear momentum $\mathbf{P}^n$ is given by 
\begin{equation}
\mathbf{P}^n\equiv \int d^3\mathbf{x}\left( \nabla \varphi _n^\alpha \right)
\left\{ \frac{\partial \mathcal{F}^n}{\partial \varphi _n^\alpha }-\nabla 
\mathbf{\cdot }\left( \frac{\partial \mathcal{F}^n}{\partial \nabla \varphi
_n^\alpha }\right) \right\} .  \label{momentum}
\end{equation}

For orbital angular momentum, consider a transformation of the fields by an
infinitesimal rotation 
\begin{equation}
\varphi _n^\alpha (\mathbf{x)\rightarrow }\varphi _n^{\alpha \prime }(%
\mathbf{x})=\varphi _n^\alpha (\mathbf{x})+\left( \delta \mathbf{\omega
\times x}\right) \mathbf{\cdot }\nabla \varphi _n^\alpha (\mathbf{x})+\delta 
\mathbf{\omega \cdot s}^{\alpha \beta }\varphi _n^\beta \left( \mathbf{x}%
\right) ,
\end{equation}
where $\mathbf{s}^{\alpha \beta }$ is the transformation matrix for the spin
indices. From this we may construct the conserved angular momentum 
\begin{equation}
\mathbf{L}^n\equiv \int d^3\mathbf{x}\left( \mathbf{x\times }\nabla \varphi
_n^\alpha +\mathbf{s}^{\alpha \beta }\varphi _n^\beta \right) \left\{ \frac{%
\partial \mathcal{F}^n}{\partial \varphi _n^\alpha }-\nabla \mathbf{\cdot }%
\left( \frac{\partial \mathcal{F}^n}{\partial \nabla \varphi _n^\alpha }%
\right) \right\} .  \label{aob}
\end{equation}

Other conserved quantities such as electric charge are just as readily
obtained by the same method.

\section{The discrete time Schr\"{o}dinger equation}

In \textit{Paper I} we discussed the possible non-commutativity of the
processes of quantisation $\mathcal{Q}$ and discretisation $\mathcal{D}$. In
that paper we presented an approach to $\mathcal{Q}$ after the application
of $\mathcal{D}$ to point particle Lagrangians. Here we shall consider
applying our $\mathcal{D}$ to a case where $\mathcal{Q}$ has already been
carried out. The first application of our approach to the temporal
discretisation of field theories is chosen to be the Schr\"{o}dinger wave
equation for a particle in an external time-independent potential.

First we note that the Schr\"{o}dinger equation may be derived as a
classical field equation in continuous time mechanics using the wave
function $\Psi $ and its complex conjugate $\Psi ^{*}$ as dynamical field
variables with the Lagrangian density 
\begin{equation}
\mathcal{L}=\frac{_1}{^2}i\hbar (\Psi ^{*}\partial _t\Psi -\partial _t\Psi
^{*}\Psi )-\frac{\hbar ^2}{2m}\Psi ^{*}\overleftarrow{\nabla }\cdot 
\overrightarrow{\nabla }\Psi -V\left( \mathbf{x}\right) \Psi ^{*}\Psi .
\end{equation}

The Euler-Lagrange equation of motion 
\begin{equation}
\partial _t\frac{\partial \mathcal{L}}{\partial \partial _t\Psi ^{*}}+\nabla
\cdot \frac{\partial \mathcal{L}}{\partial \nabla \Psi ^{*}}\stackunder{c}{=}%
\frac{\partial \mathcal{L}}{\partial \Psi ^{*}}
\end{equation}
leads to the usual Schr\"{o}dinger equation 
\begin{equation}
i\hbar \partial _t\Psi \stackunder{c}{=}-\frac{\hbar ^2\nabla ^2}{2m}\Psi
+V\left( \mathbf{x}\right) \Psi
\end{equation}
and similarly for the complex conjugate wave-function.

Turning to our discretisation process, we define our virtual paths by 
\begin{eqnarray}
\tilde{\Psi}_n\left( \tilde{t},\mathbf{x}\right)  &\equiv &\lambda \Psi
_{n+1}\left( \mathbf{x}\right) +\bar{\lambda}\Psi _n\left( \mathbf{x}\right)
,\;\;\;\;\;  \nonumber \\
\tilde{\Psi}_n^{*}\left( \tilde{t},\mathbf{x}\right)  &\equiv &\lambda \Psi
_{n+1}^{*}\left( \mathbf{x}\right) +\bar{\lambda}\Psi _n^{*}\left( \mathbf{x}%
\right) ,  \nonumber \\
\tilde{t}\left( \lambda \right)  &\equiv &t_n+\lambda \left(
t_{n+1}-t_n\right) ,
\end{eqnarray}
for $\tilde{t}\in \left[ t_n,t_{n+1}\right] $, where $\Psi _n\left( \mathbf{x%
}\right) \equiv \Psi \left( t_n,\mathbf{x}\right) ,$ etc. and $0\leq \lambda
\leq 1.$ Then the system function turns out to be 
\begin{eqnarray}
\mathcal{F}^n &=&\frac{i\hbar }2\left\{ \Psi _n^{*}\Psi _{n+1}-\Psi
_{n+1}^{*}\Psi _n\right\}   \nonumber \\
&&-\frac T6\left\{ 2\left| \Psi _{n+1}\right| ^2+2\left| \Psi _n\right|
^2+\Psi _{n+1}^{*}\Psi _n+\Psi _n^{*}\Psi _{n+1}\right\} V\left( \mathbf{x}%
\right)   \nonumber \\
&&-\frac{\hbar ^2T}{12m}\left\{ 2\left| \nabla \Psi _{n+1}\right| ^2+2\left|
\nabla \Psi _n\right| ^2+\Psi _{n+1}^{*}\overleftarrow{\nabla }\cdot 
\overrightarrow{\nabla }\Psi _n+\Psi _n^{*}\overleftarrow{\nabla }\cdot 
\overrightarrow{\nabla }\Psi _{n+1}\right\} .  \label{schro}
\end{eqnarray}

Using $\left( \ref{schro}\right) $, Cadzow's equation of motion for $\Psi $
is 
\begin{equation}
i\hbar \frac{(\Psi _{n+1}-\Psi _{n-1})}{2T}\stackunder{c}{=}\left\{ -\frac{%
\hbar ^2\nabla ^2}{2m}+V\left( \mathbf{x}\right) \right\} \frac{(\Psi
_{n+1}+4\Psi _n+\Psi _{n-1})}6  \label{s1}
\end{equation}
and similarly for the complex conjugate. This is our discretisation of the
Schr\"{o}dinger equation and represents the process $\mathcal{DQ}\,$ applied
to a standard continuous time mechanics point particle system.

Although in this theory there is no direct concept of a Hamiltonian, this
does not mean that there are no conserved quantities analogous to continuous
time energy. We define a stationary state solution to the discrete time
Schr\"{o}dinger equation $\left( \ref{s1}\right) $ to be of the form 
\begin{equation}
\Psi _n\left( \mathbf{x}\right) \equiv e^{-i\epsilon nT/\hbar }\psi \left( 
\mathbf{x}\right)  \label{state}
\end{equation}
where $\epsilon $ may be thought of as a discrete time energy. Then we find $%
\left( \ref{s1}\right) $ reduces to 
\begin{equation}
E\psi \left( \mathbf{x}\right) =\left\{ -\frac{\hbar ^2\nabla ^2}{2m}%
+V\left( \mathbf{x}\right) \right\} \psi \left( \mathbf{x}\right)
\label{conv}
\end{equation}
where the eigenvalue $E$ is given by 
\begin{equation}
E=\frac{3\hbar \sin \left( \epsilon T/\hbar \right) }{T\{\cos \left(
\epsilon T/\hbar \right) +2\}}.  \label{a33}
\end{equation}
A plot of $y\equiv ET/\hbar $ as a function of $x\equiv \epsilon T/\hbar $
shows a periodic graph which passes through $\epsilon =0$ with a slope of
unity, so that for low energies (compared to $\hbar /T$) $\epsilon $ is
virtually identical with $E$. Then the graph rises to a local maximum when 
\begin{equation}
\epsilon =\frac{2\pi \hbar }{3T}
\end{equation}
at which point $E$ has the value $E_{\max }=\frac{\sqrt{3}\hbar }T.$ If, as
we imagine, $T$ is a very small timescale, such as the Planck time, then $%
E_{\max \text{ }}$ is in practical terms far beyond any energy scale
encountered in the laboratory or in any astrophysical or cosmological
context, except possibly close to the Big Bang. Beyond the local maximum at $%
\epsilon =\frac{2\pi \hbar }{3T}$ the graph falls to zero and goes negative,
followed by a local minimum and a return to zero. Thereafter it repeats this
overall pattern. We note here the possibility of resonances, where the
values of $\epsilon $ differ by multiples of $\frac{2\pi \hbar }T.$ These
vanish in the limit $T\rightarrow $ $0$.

Assuming $0<E<\frac{\sqrt{3}\hbar }T$ and disregarding the resonances, there
are two distinct values of $\epsilon $ which would give the same value of $%
E. $ Of these, the lower value would correspond to a normal physical state
energy in practice whereas the higher value would be associated with a state
with an enormous temporal oscillation factor. We expect that such a state
would not be created in the laboratory under normal situations, because the
scattering amplitudes to go to such states oscillate very rapidly in time
and this would lead to cancellations to zero in detailed scattering
calculations.

As far as the resonances go, these give stationary solutions which differ by
at most a sign to one of the two mentioned above, and so do not represent
new states and can be ignored.

An interesting question is whether every eigenfunction of the continuous
time equation $\left( \ref{conv}\right) $ generates a stationary state
solution to $\left( \ref{s1}\right) $ of the form $\left( \ref{state}\right) 
$. The answer is no if we require $\epsilon $ to be real. For energy
eigenvalues greater than $\frac{\sqrt{3}\hbar }T,$ solutions to $\left( \ref
{a33}\right) $ will develop an imaginary component, giving discrete time
wave-functions which grow or decay exponentially, and such wave-functions
cannot be regarded as being stationary. An analogous cut-off phenomenon will
be seen when we discuss the Klein-Gordon equation in \S $5$.

To find a conserved charge we first rewrite the equation of motion (\ref{s1}%
) in the form 
\begin{eqnarray}
&&\;\;\;(3i\hbar -T\overrightarrow{S})\Psi _{n+1}\stackunder{c}{=}\left(
3i\hbar +T\overrightarrow{S}\right) \Psi _{n-1}+4T\overrightarrow{S}\Psi _n 
\nonumber \\
&&\Psi _{n+1}^{*}\left( -3i\hbar -T\overleftarrow{S}\right) \stackunder{c}{=}%
\Psi _{n-1}^{*}\left( -3i\hbar +T\overleftarrow{S}\right) +4T\Psi _n^{*}%
\overleftarrow{S}  \label{eq2}
\end{eqnarray}

where $\overrightarrow{S}\equiv \frac{-\hbar ^2}{2m}\overrightarrow{\nabla ^2%
}+V\left( \mathbf{x}\right) .$ Now consider the global gauge transformation 
\begin{equation}
\Psi _n\rightarrow e^{i\theta }\Psi _n,\;\;\;\Psi _n^{*}\rightarrow
e^{-i\theta }\Psi _n^{*}
\end{equation}
and apply the Maeda-Noether theorem discussed in \S 3. Then we find the
invariant of the motion 
\begin{equation}
Q^n\equiv \int d^3\mathbf{x}\left\{ \Psi _n^{*}(\frac{{}_1}{{}^2}+\frac{_{iT}%
}{^{6\hbar }}\overrightarrow{S})\Psi _{n+1}+\Psi _{n+1}^{*}\left( \frac{{}_1%
}{{}^2}-\overleftarrow{S}\frac{_{iT}}{^{6\hbar }}\right) \Psi _n\right\} ,
\label{inv}
\end{equation}
which reduces to the standard total probability in the limit $T\rightarrow
0.\;$With the equations of motion $\left( \ref{eq2}\right) $ we find 
\begin{equation}
Q^n\stackunder{c}{=}Q^{n-1}.
\end{equation}
From the charge $\left( \ref{inv}\right) $ we construct the charge and
current densities, given by 
\begin{eqnarray}
\rho _n &=&\Psi _n^{*}(\frac{{}_1}{{}^2}+\frac{_{iT}}{^{6\hbar }}%
\overrightarrow{S})\Psi _{n+1}+\Psi _{n+1}^{*}\left( \frac{{}_1}{{}^2}-%
\overleftarrow{S}\frac{_{iT}}{^{6\hbar }}\right) \Psi _n  \nonumber \\
\mathbf{j}_n &=&\frac{-i\hbar }{12m}\left\{ \Psi _{n+1}^{*}%
\overleftrightarrow{\nabla }\Psi _n+4\Psi _n^{*}\overleftrightarrow{\nabla }%
\Psi _n+\Psi _n^{*}\overleftrightarrow{\nabla }\Psi _{n-1}\right\} ,
\end{eqnarray}
which satisfy the discrete time equation of continuity 
\begin{equation}
\frac{\rho _n-\rho _{n-1}}T+\nabla \mathbf{\cdot j}_n\stackunder{c}{=}0.
\end{equation}

The density $\rho _n$ above is not positive definite except in the limit $%
T\rightarrow 0$, when the normal density $\Psi ^{*}\Psi $ is recovered. This
is analogous to what happens in continuous time mechanics to the charge
density for the charged Klein-Gordon equation before we take the limit $%
c\rightarrow \infty $. One of the reasons occasionally cited for
Schr\"{o}dinger's rejection of a relativistic wave equation in 1925-26 is
because of just this point, there being no evidence at that time for
positive electron charge densities. We imagine that if discrete time
mechanics had been the accepted classical mechanical paradigm prior to wave
mechanics then the Born probability interpretation of Schr\"{o}dinger wave
mechanics would probably not have been proposed.

If we wish to couple electromagnetic fields in a gauge invariant way to the
Schr\"{o}dinger equation we need to choose a modified virtual path. Such a
path is used for the charged Klein-Gordon equation discussed below, and so
this is left as an exercise in the case of the Schr\"{o}dinger equation.

\section{The Klein-Gordon field}

\subsection{\textbf{The} \textbf{continuous time system}}

An important system is the free scalar field system with Lagrange density 
\begin{equation}
\mathcal{L}\equiv \frac{{}_1}{{}^2}\partial _\mu \varphi \partial ^\mu
\varphi -\frac{{}_1}{{}^2}\mu ^2\varphi ^2,  \label{KGL}
\end{equation}
where we shall take $c=\hbar =1$ and three spatial dimensions. Hamilton's
principal function is given by 
\begin{eqnarray}
&&S^n\stackunder{c}{=}\frac{{}_1}{{}^2}\int \frac{d^3\mathbf{p}}{\left( 2\pi
\right) ^3}\frac{p_0}{\sin \left( p_0T\right) }\int d^3\mathbf{x}d^3\mathbf{%
ye}^{i\mathbf{p\cdot }\left( \mathbf{x-y}\right) }\times   \label{momq} \\
&&\;\;\;\;\;\;\;\;\;\left\{ \left[ \varphi _{n+1}\left( \mathbf{x}\right)
\varphi _{n+1}\left( \mathbf{y}\right) +\varphi _n\left( \mathbf{x}\right)
\varphi _n\left( \mathbf{y}\right) \right] \cos \left( p_0T\right) -2\varphi
_{n+1}\left( \mathbf{x}\right) \varphi _n\left( \mathbf{y}\right) \right\} ,
\nonumber
\end{eqnarray}
where $p_0\equiv \sqrt{\mathbf{p\cdot p+}\mu ^2.}$ Now compare this
principal function with that for the continuous time mechanics harmonic
oscillator, where the phenomenon of recurrence occurs, as discussed in 
\textit{Paper I}. We note that in the field theory, recurrence occurs in
momentum space. By inspection of $\left( \ref{momq}\right) ,$ singularities
in the momentum space integrand appear to arise whenever $p_0T=m\pi ,$ $%
m=1,2,...$ To understand this here we first construct the conjugate momenta
in terms of the end-point field values using $\left( \ref{aa2}\right) .$
From the principal function $\left( \ref{momq}\right) $ we find 
\begin{equation}
\pi _n\left( \mathbf{x}\right) =\int \frac{d^3\mathbf{p}}{\left( 2\pi
\right) ^3}\frac{p_0}{\sin \left( p_0T\right) }\int d^3\mathbf{y\,e}^{i%
\mathbf{p\cdot }\left( \mathbf{x-y}\right) }\left\{ \varphi _{n+1}\left( 
\mathbf{y}\right) -\varphi _n\left( \mathbf{y}\right) \cos \left(
p_0T\right) \right\} ,  \label{mom}
\end{equation}
which satisfies the expected relation 
\begin{equation}
\lim_{T\rightarrow 0}\pi _n\left( \mathbf{x}\right) =\dot{\varphi}_n\left( 
\mathbf{x}\right) 
\end{equation}
if we assume the fields vary smoothly with time.

Important constructions are the momentum fields 
\begin{eqnarray}
A\left( \mathbf{p}\right) &\equiv &\int d^3\mathbf{x\,}e^{ipx}\left\{
p_0\varphi +i\pi \right\}  \label{a1} \\
A^{*}\left( \mathbf{p}\right) &\equiv &\int d^3\mathbf{x\;}e^{-ipx}\left\{
p_0\varphi -i\pi \right\} ,  \label{a2}
\end{eqnarray}
which are time independent modulo the Klein-Gordon equation. Using $\left( 
\ref{mom}\right) $ we find 
\begin{equation}
A_n\left( \mathbf{p}\right) \equiv \frac{ip_0}{\sin \left( p_0T\right) }\int
d^3\mathbf{x\,}e^{inp_0T-i\mathbf{p\cdot x}}\left\{ \varphi _{n+1}\left( 
\mathbf{x}\right) -e^{ip_0T}\varphi _n\left( \mathbf{x}\right) \right\}
\label{a3}
\end{equation}
and similarly for its complex conjugate. The time independence of the above
momentum fields means that 
\begin{equation}
A_n\left( \mathbf{p}\right) \stackunder{c}{=}A_{n+1}\left( \mathbf{p}\right)
.  \label{const}
\end{equation}
Moreover, the momentum fields are not singular, as is evident from $\left( 
\ref{a1},\ref{a2}\right) $, so that in the limit $p_0T\rightarrow m\pi $ the
apparent singularity in $A_n\left( \mathbf{p}\right) $ must be cancelled by
recurrence behaviour in the Fourier transforms 
\begin{equation}
\tilde{\varphi}_n\left( \mathbf{p}\right) \equiv \int d^3\mathbf{x}e^{-i%
\mathbf{p\cdot x}}\varphi _n\left( \mathbf{x}\right) .
\end{equation}

We deduce from $\left( \ref{const}\right) $ that the Fourier transformed
fields $\tilde{\varphi}_n\left( \mathbf{p}\right) $ satisfy the equation of
motion 
\begin{equation}
\tilde{\varphi}_{n+1}\left( \mathbf{p}\right) +\tilde{\varphi}_{n-1}\left( 
\mathbf{p}\right) \stackunder{c}{=}2\cos \left( p_0T\right) \tilde{\varphi}%
_n\left( \mathbf{p}\right) .
\end{equation}
This shows that the continuous time Klein-Gordon field behaves like the
discrete time harmonic oscillator discussed in \textit{Paper I} and the
discrete time Klein-Gordon field discussed next, but with one important
difference. The magnitude of the factor $\cos \left( p_0T\right) $ in the
above equation of motion never exceeds unity and this ensures that the
motion is never hyperbolic. There is no need therefore for a momentum
cut-off in the continuous time theory.

\subsection{\textbf{The discrete time Klein-Gordon field}}

$~$We now consider discretising the continuous time system with Lagrange
density $\left( \ref{KGL}\right) .$ The virtual paths are of the form 
\begin{equation}
\tilde{\varphi}\left( \mathbf{x}\right) \equiv \lambda \varphi _{n+1}\left( 
\mathbf{x}\right) +\bar{\lambda}\varphi _n\left( \mathbf{x}\right)
,\;\;\;0\leq \lambda \leq 1,
\end{equation}
where $\bar{\lambda}=1-\lambda .$ With this choice of virtual path the time
derivative $\partial _t$ may be replaced by the operator $T^{-1}\partial
_\lambda ,$ which acts on $\tilde{\varphi}$ as a difference operator, viz 
\begin{equation}
\partial _t\tilde{\varphi}=\frac{\varphi _{n+1}-\varphi _n}T.
\end{equation}
The system function $F^n=F\left( \varphi _n,\varphi _{n+1},\nabla \varphi
_n,\nabla \varphi _{n+1}\right) $ is then the spatial integral 
\begin{equation}
F^n=\int d^3\mathbf{x\,}\mathcal{F}^n(\mathbf{x})
\end{equation}
of the system function density $\mathcal{F}^n(\mathbf{x)}$, given by 
\begin{equation}
\mathcal{F}^n(\mathbf{x)\equiv \,}T\int\limits_0^1d\lambda \,\mathcal{L}%
\left( \tilde{\varphi},\partial _\mu \tilde{\varphi}\right) .
\end{equation}
We find 
\begin{eqnarray}
\mathcal{F}^n &=&{\ \frac{({\varphi }_{n+1}-{\varphi }_n)^2}{2T}}-{\frac T6}%
(\left| \nabla {\varphi }_{n+1}\right| ^2+\nabla {\varphi }_{n+1}\cdot
\nabla {\varphi }_n+\left| \nabla {\varphi }_n\right| ^2)  \nonumber \\
&&-{\frac{{\mu }^2T}6}({{\varphi }_{n+1}}^2+{\varphi }_{n+1}{\varphi }_n+{{%
\varphi }_n}^2)
\end{eqnarray}
from which Cadzow's equation of motion $\left( \ref{Cadzow}\right) $ gives 
\begin{equation}
\frac{\varphi _{n+1}-2\varphi _n+\varphi _{n-1}}{T^2}+\overrightarrow{K}%
\frac{\left( \varphi _{n+1}+4\varphi _n+\varphi _{n-1}\right) }6\stackunder{c%
}{=}0
\end{equation}
where $\overrightarrow{K}\equiv -\nabla ^2+\mu ^2.$ This equation can be
written in the form 
\begin{equation}
\overrightarrow{D}\varphi _{n+1}+4\overrightarrow{D}\varphi _n+%
\overrightarrow{D}\varphi _{n-1}\stackunder{c}{=}\frac 6T\varphi _n
\label{eqm}
\end{equation}
where 
\begin{equation}
\overrightarrow{D}\equiv \frac{6+T^2\overrightarrow{K}}{6T},  \label{D}
\end{equation}
which is useful for proving various quantities are conserved. For example,
the linear momentum obtained using (\ref{momentum}) is found to be 
\begin{equation}
\mathbf{P}^n=-\int d^3\mathbf{x}\varphi _n\overleftarrow{\nabla }%
\overrightarrow{D}\varphi _{n+1}
\end{equation}
and using $\left( \ref{eqm}\right) $ we find 
\begin{equation}
\mathbf{P}^n\stackunder{c}{=}\mathbf{P}^{n-1}.
\end{equation}
Likewise, the orbital angular momentum is conserved and given by 
\begin{equation}
\mathbf{L}^n=-\int d^3\mathbf{x}\left( \mathbf{x}\times \nabla \varphi
_n\right) \overrightarrow{D}\varphi _{n+1}.
\end{equation}
These invariants of the motion exist because we have not destroyed Euclidean
invariance, that is, there is still spatial translational and rotational
invariance in our approach. This will occur in all our discrete time field
models based on special relativistic Lagrangians. We note that a similar
equation was found for the scalar field in the $\kappa -$Poincar\'{e} theory
of Lukierski et al. $\cite{LUKIERSKI.92}$, with the important differences
that their lattice parameter corresponding to our $T$ is imaginary and that
they are dealing with continuous time throughout.

We turn now to particle-like solutions to the equation of motion. Consider
the Fourier transform 
\begin{equation}
\tilde{\varphi}_n\left( \mathbf{p}\right) =\int d^3\mathbf{x\,}e^{-i\mathbf{%
p\cdot x}}\varphi _n\left( \mathbf{x}\right) .
\end{equation}
The equation of motion $\left( \ref{eqm}\right) $ becomes 
\begin{equation}
\beta \tilde{\varphi}_{n+1}+4\beta \tilde{\varphi}_n+\beta \tilde{\varphi}%
_{n-1}\stackunder{c}{=}\frac 6T\tilde{\varphi}_n,\;\;\;\;
\end{equation}
where $\;\beta \equiv \frac{6+T^2E^2}{6T}$ and $E\equiv \sqrt{\mu ^2+\mathbf{%
p\cdot p}}.$ Now define the momentum functions 
\begin{eqnarray}
a_n\left( \mathbf{p}\right) &=&i\beta e^{in\theta }\left[ \tilde{\varphi}%
_{n+1}-e^{i\theta }\tilde{\varphi}_n\right] , \\
a_n^{*}\left( \mathbf{p}\right) &=&-i\beta e^{-in\theta }\left[ \tilde{%
\varphi}_{n+1}^{*}-e^{-i\theta }\tilde{\varphi}_n^{*}\right] ,
\end{eqnarray}
where $\cos \theta =\eta ,$ $\sin \theta =\sqrt{1-\eta ^2}$ with $\eta $
given by the ratio

\begin{equation}
\eta =\frac \alpha \beta ,\,\,\,\,\,\;\;\;\;\alpha =\frac{6-2T^2E^2}{6T}.
\end{equation}
These are closely related to the continuous time momentum functions $\left( 
\ref{a3}\right) $ and to the ladder operators discussed in \textit{Paper I}
when we identify $e^{i\theta }=\mu \equiv \eta +i\sqrt{1-\eta ^2}.$ Particle
states in discrete time field theory correspond to elliptic type wave
behaviour, which requires $\left| \eta \right| <1$. Otherwise, physically
unacceptable hyperbolic behaviour occurs, as discussed in Paper \textit{I}.
This is equivalent to the condition 
\begin{equation}
TE<2\sqrt{3},  \label{limit}
\end{equation}
i.e. 
\begin{equation}
c\sqrt{\mathbf{p\cdot p}+\mu ^2c^2}<\frac{2\sqrt{3}\hbar }T
\end{equation}
in Standard International units. The fundamental conclusion from this
analysis is that energy and momentum are bounded above for physical particle
states in our approach. In\ other words, in discrete time field theory, the
spectrum of acceptable particle states has a natural cut-off which can be
made as large enough to avoid a violation of observed particle data as
necessary by choosing the time interval $T$ small enough.

An interesting point is that for a given $T,$ there can be no scalar
particle species with rest mass greater than $2\sqrt{3}\hbar /Tc^2$, but
this cannot be regarded as of interest to experimentalists at this time.

We note that by construction the momentum fields are time independent and
satisfy 
\begin{equation}
a_n\left( \mathbf{p}\right) \stackunder{c}{=}a_{n-1}\left( \mathbf{p}\right)
,\;\;\;a_n^{*}\left( \mathbf{p}\right) \stackunder{c}{=}a_{n-1}^{*}\left( 
\mathbf{p}\right) .\;
\end{equation}
This allows us to construct a Logan invariant \cite{LOGAN.73} of the motion 
\begin{equation}
C^n\equiv \int \frac{d^3\mathbf{p}}{\left( 2\pi \right) ^3}\theta \left( 
\frac{12}{T^2}-\mu ^2-\mathbf{p\cdot p}\right) a_n^{*}\left( \mathbf{p}%
\right) a_n\left( \mathbf{p}\right)
\end{equation}
which in the limit $T\rightarrow 0$ becomes the standard field theory
Hamiltonian, using the result 
\begin{equation}
\lim_{T\rightarrow 0,\;nT\rightarrow t}a_n\left( \mathbf{p}\right) =i\int d^3%
\mathbf{x}e^{iEt-i\mathbf{p\cdot x}}\left[ \dot{\varphi}(t,\mathbf{x)-}%
iE\varphi \left( t,\mathbf{x}\right) \right] =A\left( \mathbf{p}\right) .
\end{equation}
for those trajectories for which the limit exists. It is important to
understand however that there is no notion of a Hamiltonian in our approach,
since there is no possibility of infinitesimal translation in time, except
in the above limit.

\section{The free charged Klein Gordon equation}

For this system we consider the continuous time Lagrange density 
\begin{equation}
\mathcal{L}=\partial _\mu \varphi ^{*}\partial ^\mu \varphi -\mu ^2\varphi
^{*}\varphi .
\end{equation}
The virtual paths are given as before by 
\begin{equation}
\tilde{\varphi}\equiv \lambda \varphi _{n+1}+\bar{\lambda}\varphi _n,\;\;\;\;%
\tilde{\varphi}^{*}\equiv \lambda \varphi _{n+1}^{*}+\bar{\lambda}\varphi
_n^{*},\;\;\;\;\;0\leq \lambda \leq 1,
\end{equation}
where we suppress the dependence on $\mathbf{x.}$ Then the system function
density is found to be 
\begin{eqnarray}
\mathcal{F}^n &=&\frac{\left| \varphi _{n+1}-\varphi _n\right| ^2}T-\frac
T6\left\{ 2\left| \nabla \varphi _{n+1}\right| ^2+2\left| \nabla \varphi
_n\right| ^2+\nabla \varphi _{n+1}^{*}\cdot \nabla \varphi _n+\nabla \varphi
_n^{*}\cdot \nabla \varphi _{n+1}\right\}   \nonumber \\
&&-\frac{\mu ^2T}6\left\{ 2\left| \varphi _{n+1}\right| ^2+2\left| \varphi
_n\right| ^2+\varphi _{n+1}^{*}\varphi _n+\varphi _n^{*}\varphi
_{n+1}\right\} .  \label{zz1}
\end{eqnarray}
From this Cadzow's equations of motion give 
\begin{eqnarray}
&&\frac{\varphi _{n+1}-2\varphi _n+\varphi _{n-1}}{T^2}+\overrightarrow{K}%
\frac{\left( \varphi _{n+1}+4\varphi _n+\varphi _{n-1}\right) }6\stackunder{c%
}{=}0 \\
&&\frac{\varphi _{n+1}^{*}-2\varphi _n^{*}+\varphi _{n-1}^{*}}{T^2}+%
\overrightarrow{K}\frac{\left( \varphi _{n+1}^{*}+4\varphi _n^{*}+\varphi
_{n-1}^{*}\right) }6\stackunder{c}{=}0,
\end{eqnarray}
which can be written in the form 
\begin{eqnarray}
&&\overrightarrow{D}\varphi _{n+1}+4\overrightarrow{D}\varphi _n+%
\overrightarrow{D}\varphi _{n-1}\stackunder{c}{=}\frac 6T\varphi _n \\
&&\overrightarrow{D}\varphi _{n+1}^{*}+4\overrightarrow{D}\varphi _n^{*}+%
\overrightarrow{D}\varphi _{n-1}^{*}\stackunder{c}{=}\frac 6T\varphi _n^{*},
\end{eqnarray}
where $\overrightarrow{D}$ is given by (\ref{D}).

The linear and angular momenta are easy to construct so we turn to the new
feature, global gauge invariance. The system function density $\left( \ref
{zz1}\right) $ is invariant to a global gauge transformation of the fields,
i.e. 
\begin{equation}
\varphi _n\rightarrow \varphi _n^{\prime }=e^{-i\chi }\varphi
_n,\;\;\;\;\;\varphi _n^{*}\rightarrow \varphi _n^{*\prime }=e^{i\chi
}\varphi _n^{*},
\end{equation}
where $\chi $ is independent of time and space, so using the Maeda-Noether
theorem discussed in \S 3 we find the conserved charge 
\begin{eqnarray}
Q^n &\equiv &i\int d^3\mathbf{x}\left\{ \varphi _n^{*}\overrightarrow{D}%
\varphi _{n+1}-\varphi _{n+1}^{*}\overleftarrow{D}\varphi _n\right\} \\
&=&i\int d^3\mathbf{x}\left\{ \varphi _n^{*}\overleftarrow{D}\varphi
_{n+1}-\varphi _{n+1}^{*}\overrightarrow{D}\varphi _n\right\} .
\end{eqnarray}
This is real and global gauge invariant. Using the equations of motion we
find 
\begin{equation}
Q^n\stackunder{c}{=}Q^{n-1}.
\end{equation}

By inspection there are two possible candidates for a charge density,
denoted by $\rho ^{(-)}$ and $\rho ^{\left( +\right) },$ given by 
\begin{eqnarray}
\rho _n^{\left( -\right) } &\equiv &i\varphi _n^{*}\overrightarrow{D}\varphi
_{n+1}-i\varphi _{n+1}^{*}\overleftarrow{D}\varphi _n,\;\;\;\;\;\; \\
\rho _n^{\left( +\right) } &\equiv &i\varphi _n^{*}\overleftarrow{D}\varphi
_{n+1}-i\varphi _{n+1}^{*}\overrightarrow{D}\varphi _n.
\end{eqnarray}
These are related by a total divergence, that is 
\begin{equation}
\rho _n^{\left( +\right) }=\rho ^{\left( -\right) }{}_n+\frac{iT}6\nabla
\cdot \left[ \varphi _n^{*}\overleftrightarrow{\nabla }\varphi
_{n+1}+\varphi _{n+1}^{*}\overleftrightarrow{\nabla }\varphi _n\right]
\end{equation}
so that they give the same total charge $Q^n.$ Using the equations of motion
we find 
\begin{eqnarray}
&&\frac{\rho _n^{\left( -\right) }-\rho _{n-1}^{\left( -\right) }}T%
\stackunder{c}{=}\frac i6\nabla \cdot \left[ \varphi _{n-1}^{*}%
\overleftrightarrow{\nabla }\varphi _n+\varphi _n^{*}\overleftrightarrow{%
\nabla }\varphi _{n-1}+4\varphi _n^{*}\overleftrightarrow{\nabla }\varphi
_n\right] \\
&&\frac{\rho _n^{\left( +\right) }-\rho _{n-1}^{\left( +\right) }}T%
\stackunder{c}{=}\frac i6\nabla \cdot \left[ \varphi _n^{*}%
\overleftrightarrow{\nabla }\varphi _{n+1}+\varphi _{n+1}^{*}%
\overleftrightarrow{\nabla }\varphi _n+4\varphi _n^{*}\overleftrightarrow{%
\nabla }\varphi _n\right] ,
\end{eqnarray}
which are discrete time versions of the charge continuity equation, with
corresponding charge currents given by 
\begin{eqnarray}
\mathbf{j}_n^{\left( -\right) } &=&\frac{-i}6\left[ \varphi _{n-1}^{*}%
\overleftrightarrow{\nabla }\varphi _n+\varphi _n^{*}\overleftrightarrow{%
\nabla }\varphi _{n-1}+4\varphi _n^{*}\overleftrightarrow{\nabla }\varphi
_n\right] ,\;\;\; \\
\;\mathbf{j}_n^{\left( +\right) } &=&\frac{-i}6\left[ \varphi _n^{*}%
\overleftrightarrow{\nabla }\varphi _{n+1}+\varphi _{n+1}^{*}%
\overleftrightarrow{\nabla }\varphi _n+4\varphi _n^{*}\overleftrightarrow{%
\nabla }\varphi _n\right] .
\end{eqnarray}

\section{Maxwell's equations}

\subsection{Charge free equations}

We retain our unit system such that $c=\hbar =1.$ Our discrete time
formulation of charge free Maxwell's equations starts with the
electromagnetic potentials $\phi $ and $\mathbf{A}$ which are used to
construct the physical electric and magnetic fields $\mathbf{E}$ and $%
\mathbf{B}$. A clear distinction has to be made here between the nature of
the electric scalar potential $\phi $ and the magnetic vector potential $%
\mathbf{A}$. The former is associated with the \textit{temporal interval }or%
\textit{\ link }connecting times $t_n$ and $t_{n+1},$ whereas the latter is
associated with the times themselves. This distinction also manifests itself
in the difference between the physical electric field $\mathbf{E}$ and the
magnetic field $\mathbf{B}$, which are likewise associated with temporal
links and endpoints respectively.

Our definitions are as follows:

The electric scalar potential associated with the link connecting time $t_n$
and $t_{n+1}$ at spatial position $\mathbf{x}$ will be denoted by the symbol 
$\phi _n\left( \mathbf{x}\right) ,$ rather than by (say) $\phi _{n+\frac
12}\left( \mathbf{x}\right) .$ Although our notation suggests a bias towards 
$t_n$ at the expense of $t_{n+1},$ this is not really the case. The virtual
path $\tilde{\phi}_n$ for the electric potential is defined by 
\begin{equation}
\tilde{\phi}_n\left( \mathbf{x}\right) \equiv \phi _n\left( \mathbf{x}%
\right) ,\;\;\;\;0\leq \lambda \leq 1.
\end{equation}
On the other hand the magnetic vector potential associated at time $t_n$ at
spatial position $\mathbf{x}$ will be denoted by $\mathbf{A}_n\left( \mathbf{%
x}\right) $ and its associated virtual path $\mathbf{\tilde{A}}_n$ is
defined by 
\begin{equation}
\mathbf{\tilde{A}}_n\left( \mathbf{x}\right) \equiv \lambda \mathbf{A}%
_{n+1}\left( \mathbf{x}\right) +\bar{\lambda}\mathbf{A}_n\left( \mathbf{x}%
\right) ,\;\;\;0\leq \lambda \leq 1.
\end{equation}

A \textit{local gauge transformation} involves the \textit{gauge functions }$%
\chi $, which are associated with endpoints rather than links. A gauge
function value at time $t_n$ and position $\mathbf{x}$ will be denoted by $%
\chi _n\left( \mathbf{x}\right) $ and is assumed differentiable with respect
to $\mathbf{x.}$ We shall see in the next section on charged scalar field
electrodynamics that the virtual path for the gauge fields is given by 
\begin{equation}
\tilde{\chi}_n\left( \mathbf{x}\right) \equiv \lambda \chi _{n+1}\left( 
\mathbf{x}\right) +\bar{\lambda}\chi _n\left( \mathbf{x}\right) ,\;\;\;0\leq
\lambda \leq 1.
\end{equation}
A local gauge transformation is defined here by the replacements 
\begin{equation}
\mathbf{\tilde{A}}_n^{\;\mu }\rightarrow \mathbf{\tilde{A}}_n^{\prime \mu }=%
\mathbf{\tilde{A}}_n^{\;\mu }+\partial ^\mu \tilde{\chi}_n,
\end{equation}
which reduce to the transformations 
\begin{eqnarray}
\phi _n &\rightarrow &\phi _n^{\prime }=\phi _n+\frac{\chi _{n+1}-\chi _n}T,
\label{gauge1} \\
\mathbf{A}_n &\rightarrow &\mathbf{A}_n^{\prime }=\mathbf{A}_n-\nabla \chi
_n.  \label{gauge2}
\end{eqnarray}

Turning now to the physical fields, we define the gauge invariant electric
and magnetic fields via the potentials: 
\begin{equation}
\mathbf{E}_n\mathbf{\equiv -}\nabla \phi _n-\frac{\left( \mathbf{A}_{n+1}-%
\mathbf{A}_n\right) }T,\;\;\;\;\;\mathbf{B}_n\equiv \nabla \times \mathbf{A}%
_n,
\end{equation}
which give the homogeneous Maxwell equations 
\begin{equation}
\nabla \cdot \mathbf{B}_n=0,\;\;\;\;\;\nabla \times \mathbf{E}_n+\frac{%
\left( \mathbf{B}_{n+1}-\mathbf{B}_n\right) }T=\mathbf{0.}  \label{Maxwell1}
\end{equation}

The gauge invariant system function density for the charge free system is
defined as the integral 
\begin{equation}
\mathcal{F}^n\equiv T\int\limits_0^1d\lambda \left( -\frac{_1}{^4}\tilde{F}%
_{n\mu \nu }\tilde{F}_n^{\;\mu \nu }\right) ,
\end{equation}
where $\tilde{F}_n^{\;\mu \nu }\equiv $ $\partial ^\mu \tilde{A}_n^{\;\nu
}-\partial ^\nu \tilde{A}_n^{\;\mu }$ are the components of the Faraday
tensor. This gives \textbf{\ } 
\begin{equation}
\mathcal{F}^n=\frac T2\mathbf{E}_n^2-\frac T6\left( \mathbf{B}_{n+1}^2+%
\mathbf{B}_{n+1}\cdot \mathbf{B}_n+\mathbf{B}_n^2\right) .  \label{s12}
\end{equation}
Applying Cadzow's equation to $\left( \ref{s12}\right) $ we find 
\begin{equation}
\nabla \cdot \mathbf{E}_n\stackunder{c}{=}0,\;\;\;\;\;\frac{\left( \mathbf{E}%
_n-\mathbf{E}_{n-1}\right) }T\stackunder{c}{=}\nabla \times \frac{\left( 
\mathbf{B}_{n+1}+4\mathbf{B}_n+\mathbf{B}_{n-1}\right) }6.  \label{Maxwell2}
\end{equation}
Using $\left( \ref{Maxwell1}\right) $ and $\left( \ref{Maxwell2}\right) $%
\textbf{\ }we find the physical electromagnetic fields satisfy the discrete
time massless Klein-Gordon equations 
\begin{eqnarray}
&&\frac{\mathbf{E}_{n+1}-2\mathbf{E}_n+\mathbf{E}_{n-1}}{T^2}-\nabla ^2\frac{%
\left( \mathbf{E}_{n+1}+4\mathbf{E}_n+\mathbf{E}_{n-1}\right) }6\stackunder{c%
}{=}\mathbf{0,} \\
&&\frac{\mathbf{B}_{n+1}-2\mathbf{B}_n+\mathbf{B}_{n-1}}{T^2}-\nabla ^2\frac{%
\left( \mathbf{B}_{n+1}+4\mathbf{B}_n+\mathbf{B}_{n-1}\right) }6\stackunder{c%
}{=}\mathbf{0.}
\end{eqnarray}

Turning to the gauge potentials, we define the discrete time Lorentz\ gauge\
by the condition 
\begin{equation}
\frac{\phi _n-\phi _{n-1}}T+\nabla \cdot \frac{\left( \mathbf{A}_{n+1}+4%
\mathbf{A}_n+\mathbf{A}_{n-1}\right) }6=0  \label{lor}
\end{equation}
and then the gauge potentials also satisfy the discrete time massless
Klein-Gordon equations 
\begin{eqnarray}
&&\frac{\mathbf{A}_{n+1}-2\mathbf{A}_n+\mathbf{A}_{n-1}}{T^2}-\nabla ^2\frac{%
\left( \mathbf{A}_{n+1}+4\mathbf{A}_n+\mathbf{A}_{n-1}\right) }6\stackunder{c%
}{=}\mathbf{0,} \\
&&\;\;\;\frac{\phi _{n+1}-2\phi _n+\phi _{n-1}}{T^2}-\nabla ^2\frac{\left(
\phi _{n+1}+4\phi _n+\phi _{n-1}\right) }6\stackunder{c}{=}0.
\end{eqnarray}

The total linear momentum for the free electromagnetic fields is found to be 
\begin{equation}
\mathbf{P}^n=\int d^3\mathbf{x}\left\{ \mathbf{E}_n\mathbf{\times B}_n+\frac
T6B_n^i\overrightarrow{\nabla }B_{n+1}^i\right\} .
\end{equation}
Then using the equations of motion we find 
\begin{equation}
\mathbf{P}^n\stackunder{c}{=}\mathbf{P}^{n-1}
\end{equation}
as expected. In the limit $T\rightarrow 0$ the above expression reduces to
the Poynting vector.

\subsection{Maxwell's equations in the presence of charges}

In the presence of electric charges the system function density $\left( \ref
{s12}\right) $ is replaced by 
\begin{equation}
\mathcal{F}^n=\frac T2\mathbf{E}_n^2-\frac T6\left( \mathbf{B}_{n+1}^2+%
\mathbf{B}_{n+1}\cdot \mathbf{B}_n+\mathbf{B}_n^2\right) -T\phi _n\rho _n+T%
\mathbf{A}_n\cdot \mathbf{j}_n
\end{equation}
where $\rho _n\left( \mathbf{x}\right) $ and $\mathbf{j}_n\left( \mathbf{x}%
\right) $ are the discrete time charge density and charge current
respectively. The homogeneous equations $\left( \ref{Maxwell1}\right) $
remain unaltered but now the equations of motion become 
\begin{equation}
\nabla \cdot \mathbf{E}_n\stackunder{c}{=}\rho _n,\;\;\;\;\nabla \times 
\frac{\left( \mathbf{B}_{n+1}+4\mathbf{B}_n+\mathbf{B}_{n-1}\right) }6-\frac{%
\left( \mathbf{E}_n-\mathbf{E}_{n-1}\right) }T\stackunder{c}{=}\mathbf{j}_n.
\end{equation}
These equations are consistent with the equation of continuity for electric
charge given by 
\begin{equation}
\frac{\rho _n-\rho _{n-1}}T+\nabla \cdot \mathbf{j}_n\stackunder{c}{=}0.
\end{equation}

Finally, in a discrete time Lorentz gauge $\left( \ref{lor}\right) $ the
potentials satisfy the equations 
\begin{eqnarray}
&&\frac{\mathbf{A}_{n+1}-2\mathbf{A}_n+\mathbf{A}_{n-1}}{T^2}-\nabla ^2\frac{%
\left( \mathbf{A}_{n+1}+4\mathbf{A}_n+\mathbf{A}_{n-1}\right) }6\stackunder{c%
}{=}\mathbf{j}_n,  \nonumber \\
&&\;\;\;\frac{\phi _{n+1}-2\phi _n+\phi _{n-1}}{T^2}-\nabla ^2\frac{\left(
\phi _{n+1}+4\phi _n+\phi _{n-1}\right) }6\stackunder{c}{=}\frac{\left( \rho
_{n+1}+4\rho _n+\rho _{n-1}\right) }6.  \nonumber \\
&&
\end{eqnarray}

\section{Scalar field electrodynamics}

We are now in a position to discuss the coupling of the Maxwell potentials
to the charged Klein-Gordon field. In this case the virtual paths for the
charged scalar field and its complex conjugate have to be modified so as to
ensure that the system function is gauge invariant. We define 
\begin{eqnarray}
\tilde{\varphi} &\equiv &\lambda e^{iq\phi _nT\bar{\lambda}}\varphi _{n+1}+%
\bar{\lambda}e^{-iq\phi _nT\lambda }\varphi _n,  \nonumber \\
\;\;\tilde{\varphi}^{*} &\equiv &\lambda e^{-iq\phi _nT\bar{\lambda}}\varphi
_{n+1}^{*}+\bar{\lambda}e^{iq\phi _nT\lambda }\varphi _n^{*},\;\;\;\;\;\bar{%
\lambda}=1-\lambda :\;\;\;0\leq \lambda \leq 1.  \label{virt}
\end{eqnarray}
where $\phi _n$ is the scalar potential discussed in the previous section.
Then the gauge transformations are given by ($\ref{gauge1},\ref{gauge2})$
and by 
\begin{equation}
\varphi _n^{\prime }=e^{-iq\chi _n}\varphi _n,\;\;\;\;\varphi _{n+1}^{\prime
}=e^{-iq\chi _{n+1}}\varphi _{n+1},
\end{equation}
which leads to the compact form \textbf{\ } 
\begin{equation}
\tilde{\varphi}^{\prime }=e^{-iq\tilde{\chi}}\tilde{\varphi},\;\;\;\;\tilde{%
\varphi}^{*\prime }=e^{iq\tilde{\chi}}\tilde{\varphi}^{*}.
\end{equation}

Now define the gauge covariant derivatives 
\begin{equation}
\tilde{D}_\mu \tilde{\varphi}\equiv \left( \partial _\mu +iq\tilde{A}_\mu
\right) \tilde{\varphi}
\end{equation}
and its complex conjugate. Under a gauge transformation we find 
\begin{equation}
\left( \tilde{D}_\mu \tilde{\varphi}\right) ^{\prime }\equiv e^{-iq\tilde{%
\chi}}\tilde{D}_\mu \tilde{\varphi}.
\end{equation}
Next we construct the gauge invariant Lagrange density 
\begin{eqnarray}
\mathcal{L}^n &=&\left( \tilde{D}_\mu \tilde{\varphi}\right) ^{*}\tilde{D}%
^\mu \tilde{\varphi}-\mu ^2\tilde{\varphi}^{*}\tilde{\varphi}  \nonumber \\
&=&\left( \tilde{D}_0\tilde{\varphi}\right) ^{*}\tilde{D}^0\tilde{\varphi}%
-\left( \mathbf{\tilde{D}}\tilde{\varphi}\right) ^{*}\mathbf{\cdot \tilde{D}}%
\tilde{\varphi}-\mu ^2\tilde{\varphi}^{*}\tilde{\varphi},
\end{eqnarray}
where 
\begin{equation}
\tilde{D}_0\tilde{\varphi}\equiv \left( \frac 1T\partial _\lambda +iq\phi
_n\right) \tilde{\varphi}=\frac{e^{iq\phi _nT\bar{\lambda}}\varphi
_{n+1}-e^{-iq\phi _nT\lambda }\varphi _n}T
\end{equation}

and 
\begin{eqnarray}
\mathbf{\tilde{D}}\tilde{\varphi} &\equiv &(\nabla -iq\mathbf{\tilde{A})}%
\tilde{\varphi}  \nonumber \\
&=&iqT\lambda \bar{\lambda}\nabla \phi _n\left\{ e^{iq\phi _nT\bar{\lambda}%
}\varphi _{n+1}-e^{-iq\phi _nT\lambda }\varphi _n\right\}  \nonumber \\
&&+\lambda e^{iq\phi _nT\bar{\lambda}}\nabla \varphi _{n+1}+\bar{\lambda}%
e^{-iq\phi _nT\lambda }\nabla \varphi _n-iq\mathbf{\tilde{A}}\tilde{\varphi}.
\end{eqnarray}

The system function is constructed as before from the integral 
\begin{equation}
\mathcal{F}^n=T\int\limits_0^1d\lambda \left\{ \left( \tilde{D}_0\tilde{%
\varphi}\right) ^{*}\tilde{D}_0\tilde{\varphi}-\left( \mathbf{\tilde{D}}%
\tilde{\varphi}\right) ^{*}\mathbf{\cdot \tilde{D}}\tilde{\varphi}-\mu ^2%
\tilde{\varphi}^{*}\tilde{\varphi}\right\} ,  \label{form}
\end{equation}
which gives 
\begin{eqnarray}
\mathcal{F}^n &=&\left( \frac 1T-\frac{q^2T^3}{30}\nabla \phi _n\cdot \nabla
\phi _n\right) \left| U_n\varphi _{n+1}-\varphi _n\right| ^2  \nonumber \\
&&-\frac T6\left\{ 2\left| \nabla \varphi _{n+1}\right| ^2+2\left| \nabla
\varphi _n\right| ^2+U_n^{*}\nabla \varphi _{n+1}^{*}\cdot \nabla \varphi
_n+\nabla \varphi _n^{*}\cdot \nabla \varphi _{n+1}U_n\right\}  \nonumber \\
&&\_\frac{\mu ^2T}6\left\{ 2\left| \varphi _{n+1}\right| ^2+2\left| \varphi
_n\right| ^2+U_n^{*}\varphi _{n+1}^{*}\varphi _n+\varphi _n^{*}\varphi
_{n+1}U_n\right\}  \nonumber \\
&&+\frac{iqT^2}{12}\left\{ 
\begin{array}{c}
\left( U_n^{*}\varphi _{n+1}^{*}-\varphi _n^{*}\right) \nabla \phi _n\cdot
\left( U_n\nabla \varphi _{n+1}+\nabla \varphi _n\right) \\ 
-\nabla \phi _n\cdot \left( U_n^{*}\nabla \varphi _{n+1}^{*}+\nabla \varphi
_n^{*}\right) \left( U_n\varphi _{n+1}-\varphi _n\right)
\end{array}
\right\}  \nonumber \\
&&+\frac{q^2T^2}{60}\left\{ \left( U_n^{*}\varphi _{n+1}^{*}-\varphi
_n^{*}\right) \nabla \varphi _n\cdot \left[ \left( 3\mathbf{A}_{n+1}+2%
\mathbf{A}_n\right) U_n\varphi _{n+1}+\left( 2\mathbf{A}_{n+1}+3\mathbf{A}%
_n\right) \varphi _n\right] \right.  \nonumber \\
&&+\left. \left[ \left( 3\mathbf{A}_{n+1}+2\mathbf{A}_n\right)
U_n^{*}\varphi _{n+1}^{*}+\left( 2\mathbf{A}_{n+1}+3\mathbf{A}_n\right)
\varphi _n^{*}\right] \cdot \nabla \varphi _n\left( U_n\varphi
_{n+1}-\varphi _n\right) \right\}  \nonumber \\
&&-\frac{iqT}{12}\left\{ 
\begin{array}{c}
\left( 3\mathbf{A}_{n+1}+\mathbf{A}_n\right) \cdot \left( \varphi
_{n+1}^{*}\nabla \varphi _{n+1}-\nabla \varphi _{n+1}^{*}\varphi
_{n+1}\right) \\ 
+\left( 3\mathbf{A}_n+\mathbf{A}_{n+1}\right) \cdot \left( \varphi
_n^{*}\nabla \varphi _n-\nabla \varphi _n^{*}\varphi _n\right)
\end{array}
\right.  \nonumber \\
&&+\left. \left( \mathbf{A}_{n+1}+\mathbf{A}_n\right) \cdot \left[ \varphi
_n^{*}\nabla \varphi _{n+1}U_n-\nabla \varphi _n^{*}\varphi
_{n+1}U_n+U_n^{*}\varphi _{n+1}^{*}\nabla \varphi _n-U_n^{*}\nabla \varphi
_{n+1}^{*}\varphi _n\right] \right\}  \nonumber \\
&&-\frac{q^2T}{60}\left\{ 
\begin{array}{c}
\left( 12\mathbf{A}_{n+1}^2+2\mathbf{A}_n^2+6\mathbf{A}_n\cdot \mathbf{A}%
_{n+1}\right) \left| \varphi _{n+1}\right| ^2 \\ 
+\left( 12\mathbf{A}_n^2+2\mathbf{A}_{n+1}^2+6\mathbf{A}_n\cdot \mathbf{A}%
_{n+1}\right) \left| \varphi _n\right| ^2
\end{array}
\right.  \nonumber \\
&&+\left. \left( 3\mathbf{A}_{n+1}^2+3\mathbf{A}_n^2+4\mathbf{A}_n\cdot 
\mathbf{A}_{n+1}\right) \left[ U_n^{*}\varphi _{n+1}^{*}\varphi _n+\varphi
_n^{*}\varphi _{n+1}U_n\right] \right\}
\end{eqnarray}
where $U_n\equiv e^{iq\phi _nT}.$ This system function is gauge invariant.

The application of the resulting Cadzow's equations to specific problems is
left for possible future investigation. We make the following comments in
passing. First, this system function looks complicated. However, that is a
matter of notation, as the expression $\left( \ref{form}\right) $ is
equivalent and much more compact. There is undoubtedly much dynamical
content in this system function and calculations would no doubt require
approximations to be made. However we recall here Gell-Mann's dictum that it
is better to make approximations to exact equations than to solve exactly
approximate equations. Also, most applications would themselves be better
considered from a second quantised approach, which we shall study in \textit{%
Paper IV} of this series.

We may simplify the system function considerably by considering the special
case when the external magnetic potential $\mathbf{A}_n$ vanishes. Then the
system function reduces to 
\begin{eqnarray}
\mathcal{F}^n &=&\left( \frac 1T-\frac{q^2T^3}{30}\nabla \phi _n\cdot \nabla
\phi _n\right) \left| U_n\varphi _{n+1}-\varphi _n\right| ^2  \nonumber \\
&&-\frac T6\left\{ 2\left| \nabla \varphi _{n+1}\right| ^2+2\left| \nabla
\varphi _n\right| ^2+U_n^{*}\nabla \varphi _{n+1}^{*}\cdot \nabla \varphi
_n+\nabla \varphi _n^{*}\cdot \nabla \varphi _{n+1}U_n\right\}  \nonumber \\
&&\_\frac{\mu ^2T}6\left\{ 2\left| \varphi _{n+1}\right| ^2+2\left| \varphi
_n\right| ^2+U_n^{*}\varphi _{n+1}^{*}\varphi _n+\varphi _n^{*}\varphi
_{n+1}U_n\right\}  \nonumber \\
&&+\frac{iqT^2}{12}\left\{ 
\begin{array}{c}
\left( U_n^{*}\varphi _{n+1}^{*}-\varphi _n^{*}\right) \nabla \phi _n\cdot
\left( U_n\nabla \varphi _{n+1}+\nabla \varphi _n\right) \\ 
-\nabla \phi _n\cdot \left( U_n^{*}\nabla \varphi _{n+1}^{*}+\nabla \varphi
_n^{*}\right) \left( U_n\varphi _{n+1}-\varphi _n\right)
\end{array}
,\right\}
\end{eqnarray}
which leads to the equation of motion 
\begin{eqnarray}
&&\frac{U_n\varphi _{n+1}-2\varphi _n+U_{n-1}^{*}\varphi _{n-1}}{T^2}+\frac{%
\mu ^2\left( U_n\varphi _{n+1}+4\varphi _n+U_{n-1}^{*}\varphi _{n-1}\right) }%
6  \nonumber \\
&&-\nabla \cdot \left[ \frac{\nabla \varphi _{n+1}U_n+4\nabla \varphi
_n+U_{n-1}^{*}\nabla \varphi _{n-1}}6\right] \stackunder{c}{=}  \nonumber \\
&&\frac{q^2T^2}{30}\left\{ \left( \nabla \phi _n\right) ^2\left[ U_n\varphi
_{n+1}-\varphi _n\right] -\left( \nabla \phi _{n-1}\right) ^2\left[ \varphi
_n-U_{n-1}^{*}\varphi _{n-1}\right] \right\}  \nonumber \\
&&-\frac{iqT}{12}\left\{ \nabla \phi _n\cdot (U_n\nabla \varphi
_{n+1}+\nabla \varphi _n)-\nabla \phi _{n-1}\cdot \left( \nabla \varphi
_n+U_{n-1}^{*}\nabla \varphi _{n-1}\right) \right\}  \nonumber \\
&&+\frac{iqT}{12}\nabla \cdot \left\{ \nabla \phi _n\left( U_n\varphi
_{n+1}-\varphi _n\right) +\nabla \phi _{n-1}\left( \varphi
_n-U_{n-1}^{*}\varphi _{n-1}\right) \right\} .
\end{eqnarray}
In the limit $T\rightarrow 0$ we recover the usual equation 
\begin{equation}
\left( \partial _t+iq\phi \right) \left( \partial _t+iq\phi \right) \varphi
+\left( \mu ^2-\nabla ^2\right) \varphi \stackunder{c}{=}0.
\end{equation}

\section{The Dirac equation}

We turn now to the continuous time Dirac equation 
\begin{equation}
i\dot{\psi}\stackunder{c}{=}\left( -i\mathbf{\alpha \cdot \overrightarrow{%
\nabla }}+\beta m\right) \psi ,  \label{Dirac}
\end{equation}
which is obtained from the Lagrange density 
\begin{equation}
\mathcal{L}=\frac{_1}{^2}\psi ^{+}\left[ i\dot{\psi}+i\mathbf{\alpha \cdot 
\overrightarrow{\nabla }}\psi \right] +\frac{_1}{^2}\left[ -i\dot{\psi}%
^{+}-i\psi ^{+}\mathbf{\alpha \cdot \overleftarrow{\nabla }}\right] \psi
-m\psi ^{+}\beta \psi .
\end{equation}
The virtual paths are taken to be 
\begin{equation}
\tilde{\psi}=\lambda \psi _{n+1}+\bar{\lambda}\psi _n,\;\;\;\;\;\tilde{\psi}%
^{+}=\lambda \psi _{n+1}^{+}+\bar{\lambda}\psi _n^{+}
\end{equation}
and then we find the system function density 
\begin{eqnarray}
\mathcal{F}^n &=&\frac i2\left\{ \psi _n^{+}\psi _{n+1}-\psi _{n+1}^{+}\psi
_n\right\}   \nonumber \\
&&-\frac T6\left\{ 2\psi _{n+1}^{+}\overleftrightarrow{D}\psi _{n+1}+\psi
_{n+1}^{+}\overleftrightarrow{D}\psi _n+\psi _n^{+}\overleftrightarrow{D}%
\psi _{n+1}+2\psi _n^{+}\overleftrightarrow{D}\psi _n\right\}   \label{Dsys}
\end{eqnarray}
where $\overleftrightarrow{D}\equiv -\frac{_1}{^2}i\mathbf{\alpha \cdot 
\overleftrightarrow{\nabla }}+\beta m.$ From this we find the Cadzow
equation of motion 
\begin{equation}
i\frac{\left( \psi _{n+1}-\psi _{n-1}\right) }{2T}\stackunder{c}{=}%
\overrightarrow{H_D}\frac{\left( \psi _{n+1}+4\psi _n+\psi _{n-1}\right) }6,
\label{DDirac}
\end{equation}
where $\overrightarrow{H_D}\equiv -i\mathbf{\alpha \cdot \overrightarrow{%
\nabla }}+\beta m$ and similarly for the conjugate field $\psi _n^{+}.$ This
is the required discretisation of the Dirac equation $\left( \ref{Dirac}%
\right) .$

This equation has the same structure as that for the Grassmannian oscillator
discussed in \textit{Paper I}. To determine the regime for elliptic
(particle like) behaviour, we first Fourier transform $\left( \ref{DDirac}%
\right) $, defining 
\begin{equation}
\int d^3\mathbf{xe}^{-i\mathbf{p\cdot x}}\psi _n(\mathbf{x)=\;}\tilde{\psi}%
_n\left( \mathbf{p}\right) ,
\end{equation}
and then $\left( \ref{DDirac}\right) $ becomes 
\begin{equation}
\left( -3i+DT\right) \tilde{\psi}_{n+1}+\left( 3i+DT\right) \tilde{\psi}%
_{n-1}\stackunder{c}{=}-4DT\tilde{\psi}_n,
\end{equation}
where $D\equiv \mathbf{\alpha \cdot p}+\beta m$, which has the property 
\begin{equation}
DD\equiv \mathbf{p\cdot p}+m^2=E^2,
\end{equation}
$E$ being the definition of the particle's energy (keeping in mind there is
no 'Hamiltonian' in this theory).

Now define the Dirac matrix 
\begin{equation}
\nu \equiv \frac{3iDT+E^2T^2}{\sqrt{9E^2T^2+E^4T^4}}
\end{equation}
and rescale the momentum fields via the rule 
\begin{equation}
\tilde{\psi}_n\equiv \nu ^n\phi _n.
\end{equation}
Then we find 
\begin{equation}
\phi _{n+1}+\phi _{n-1}\stackunder{c}{=}2\eta \phi _n
\end{equation}
where $\eta =\frac{-2ET}{\sqrt{9+E^2T^2}}$, which proves that oscillator
behaviour is inherent in this system. Particle like solutions will occur for 
$\eta ^2<1$, from which we deduce the condition $ET<\sqrt{3}.$ This upper
limit is one half of the upper limit $\left( \ref{limit}\right) $
established for the discretised Klein-Gordon equation. Exactly the same
phenomenon occurs in the Grassmannian oscillator studied in \textit{Paper I}.

We shall leave the construction of conserved quantities such as the linear
momentum and the Logan invariant built up from the ladder operators $a\left( 
\mathbf{p}\right) ,b\left( \mathbf{p}\right) ,$ etc., as exercises. However,
we will discuss here the charge and charge current densities.

First, we rewrite the Cadzow equations of motion $\left( \ref{DDirac}\right) 
$ in the more useful form 
\begin{eqnarray}
&&\left( -3i+T\overrightarrow{H_D}\right) \psi _{n+1}\stackunder{c}{=}-4T%
\overrightarrow{H_D}\psi _n-\left( 3i+T\overrightarrow{H_D}\right) \psi
_{n-1}  \nonumber \\
&&\psi _{n+1}^{+}\left( 3i+\overleftarrow{H_D}T\right) \stackunder{c}{=}%
-4\psi _n^{+}\overleftarrow{H_D}T-\psi _{n-1}^{+}\left( -3i+\overleftarrow{%
H_D}T\right) ,  \label{DDeq}
\end{eqnarray}
where $\overrightarrow{H_D}\equiv -i\mathbf{\alpha \cdot \overrightarrow{%
\nabla }}+\beta m,\;\;\;\overleftarrow{H_D}\equiv i\mathbf{\alpha \cdot 
\overleftarrow{\nabla }}+\beta m$.

Now the system function $\left( \ref{Dsys}\right) $ is invariant to the
infinitesimal global gauge transformation 
\begin{eqnarray}
\psi _n &\rightarrow &e^{i\theta }\psi _n,\;\;\;\psi _n^{+}\rightarrow \psi
_n^{+}e^{-i\theta }  \nonumber \\
\delta \psi _n &=&i\theta \psi _n,\;\;\;\;\delta \psi ^{+}=-i\theta \psi
_n^{+},
\end{eqnarray}
so that there is a conserved charge given by 
\begin{equation}
Q^n\equiv \int d^3\mathbf{x}\left\{ \psi _n^{+}\left( \frac{_1}{^2}+\frac{iT}%
6\overrightarrow{H_D}\right) \psi _{n+1}+\psi _{n+1}^{+}\left( \frac{_1}{^2}-%
\overleftarrow{H_D}\frac T6\right) \psi _n\right\} .
\end{equation}
It is easy to use the equations of motion $\left( \ref{DDeq}\right) $ to
show that 
\begin{equation}
Q^n\stackunder{c}{=}Q^{n-1}.
\end{equation}
The charge and current densities are then given by 
\begin{eqnarray}
\rho _n &=&\psi _n^{+}\left( \frac{_1}{^2}+\frac{_{iT}}{^6}\overrightarrow{%
H_D}\right) \psi _{n+1}+\psi _{n+1}^{+}\left( \frac{_1}{^2}-\overleftarrow{%
H_D}\frac{_{iT}}{^6}\right) \psi _n  \nonumber \\
\mathbf{j}_n &=&\frac{_1}{^6}\left\{ \psi _n^{+}\mathbf{\alpha }\psi
_{n-1}+4\psi _n^{+}\mathbf{\alpha }\psi _n+\psi _{n-1}^{+}\mathbf{\alpha }%
\psi _n\right\} .
\end{eqnarray}
These collapse to the standard densities in the limit $T\rightarrow 0,$ and
satisfy the discrete time equation of continuity 
\begin{equation}
\frac{\rho _n-\rho _{n-1}}T+\nabla \mathbf{\cdot j}_n\stackunder{c}{=}0
\end{equation}
as required.

This completes our discussion of the free Dirac equation. The coupling of
the Dirac field to the electromagnetic field in a gauge invariant way
follows exactly the same procedure as with the charged Klein-Gordon field
discussed in \S 8, with virtual paths following the rules given by $\left( 
\ref{virt}\right) $. Second quantisation of the free Dirac field and QED
will be discussed in the fourth in this series.

\section{Conclusions}

In this paper we have extended the methods outlined in the first paper of
this series to classical fields, including scalar, vector, and spinor
fields. The construction of equations of motion and invariants of the motion
is straightforward and gauge invariance also can be readily built into the
system function.

It is clear that the discretisation process does separate out timelike and
spacelike terms in most equations, so that some of the elegance of a fully
Lorentz covariant theory is lost. This is an inevitability in any
discretisation process. However, this disadvantage should be viewed
alongside the potential advantages, which may be considerable.

First, our approach is exact, in that we do not countenance approximations
within the theory itself, only in applications. Equations of motion are
derived exactly from well-specified system functions, and invariants of the
motion are precisely that, i.e., exact invariants of the motion and not
approximate invariants of the motion. Next, there is a natural scale
introduced into the theory, namely the fundamental time interval $T$. This
influences the dynamics of the fields and leads to an upper limit for
particle state energies, where the definition of particle state requires
field amplitudes to remain bounded in time (corresponding to the condition $%
\eta ^2<1$ in the harmonic oscillator). This holds some promise as a
regulariser in quantum field theory, where in conventional calculations ad
hoc cutoffs have to be introduced routinely.

One of the surprises discrete time mechanics has sprung is how well it can
work. Although it is certainly not equivalent to continuous time mechanics,
in almost every aspect of the latter the former can make a comparable
statement. We do not have a Hamiltonian in discrete time mechanics, but we
do have a system function which can be used to generate equations of the
motion, invariants of the motion, and define quantisation, as discussed in
Paper \textit{I}. In the case of free fields we can readily construct a
Logan invariant which substitutes for the Hamiltonian in a particularly
useful way, closed allied with the particle content of the theory.

This raises the important question: if indeed all continuous time dynamical
theories can be simulated by discrete time analogues, given a small enough $%
T $, so that no empirical test could distinguish between their predictions,
what would force us to pick one approach rather than the other? One theory
is elegant but badly defined mathematically in many places, whilst the other
looks clumsier but is perhaps better defined mathematically.

Ultimately it may come down to a question of personal choice, which will be
influenced by some intrinsic features of the approach. For instance, the
construction of Feynman path integrals as infinite products of integrals
cannot be carried out in the usual way in discrete time field theory, simply
because in that theory there is no meaning to the taking of the limit $%
T\rightarrow 0$, other than to show some sort of consistency with
conventional theory. Therefore, one of the serious objections to the
conventional path integral formalism, that of being ill-defined, is greatly
moderated in discrete time theory. There is something like a path integral,
but it involves a finite number of time steps between initial and final
times, as discussed in Paper \textit{I}. In Paper $III$ of this series we
shall discuss the second quantisation of field theories using this approach.

\section{Acknowledgements}

We are pleased to thank Prof. J. Lukierski for his invaluable comments at an
early stage in this work and for material assistance in finding important
references and opportunities to discuss discrete time mechanics.

\pagebreak

\end{document}